 \documentclass[final]{elsarticle}

\usepackage{amssymb}
\usepackage{color}

\begin{document}
\newcommand{\req}[1]{(\ref{#1})}
\newcommand{\bel}[1]{\begin{equation}\label{#1}}
\newcommand{\belar}[1]{\begin{eqnarray}\label{#1}}
\def\r{{\bf r}}
\def\p{{\bf p}}\def\eps{\epsilon}
\def\mev{\;{\rm MeV}}
\def\tc{\textcolor{red}}
\def\tb{\textcolor{blue}}
\def\pr{\prime}
\def\dpr{\prime\prime}
\begin{frontmatter}




\title{Dynamic effects of nuclear surface in isoscalar dipole modes}


\author[kinr]{ V. I. Abrosimov\corref{cor1}}
\ead{abrosim@kinr.kiev.ua}
\author[kinr]{O. I. Davydovska}
\cortext[cor1]{Corresponding author}
\address[kinr]{Institute for Nuclear Research, Prospect Nauki 47, 03028 Kyiv, Ukraine}

\thispagestyle{empty}

\begin{abstract}

Dynamic surface effects in the isoscalar dipole modes of heavy nuclei are studied within a semiclassical model based on the solution of the Vlasov kinetic equation for finite Fermi systems with a moving surface.
In order to clarify the role of dynamic surface effects we have considered an approximate solution, which takes into account only the part of the variation of the phase-space distribution function 
caused by the dynamic surface (the dynamic-surface approximation).
It is shown that the dynamic surface effects have an essential influence on the features of the isoscalar dipole modes.
The isoscalar dipole strength function has a two-resonance structure already in the dynamic-surface approximation, and the centroid energies of both the low-energy resonance and the high-energy resonance are close to corresponding centroid energies of exact strength function.
Calculations of the velocity field in the dynamic-surface approximation show the vortex character of the low-energy isoscalar dipole resonance and the compression character of the high-energy one. 
\end{abstract}



\end{frontmatter}


\section{Introduction}
\label{intro}

Experimental studies of the isoscalar dipole nuclear response have shown not only the high-energy resonance related to the compressibility of nuclear matter (the compression mode), but also the low-energy resonance (the vortex mode) 
\cite{Clark-01}-\cite{Uchida-04}.
Various theoretical studies of nuclear isoscalar dipole excitations have been carried out within quantum approaches of the random phase approximation (RPA) type, including their relativistic generalization 
\cite{Giai-81}-\cite{Kvasil-14}, 
as well as within semiclassical approaches based on the phase-space dynamics 
\cite{Balb-94}-\cite{Abr-18}.
It is known that isoscalar dipole excitations of finite many-body systems are related to the center-of -mass motion of the system and therefore, in a theoretical study of the isoscalar dipole excitations in nuclei,
the motion of the center of mass should be separated from internal isoscalar dipole excitations. Moreover, the isoscalar dipole excitations are considered as the second-order effect in the dipole moment because in the first order they are reduced to the center-of-mass motion of the system. 
Taking into account these features of nuclear isoscalar dipole excitations, one can expect a significant influence of surface effects on the dynamics of isoscalar dipole excitations, and therefore it is of interest to study this problem in more detail.

In this paper, we study the effect of  dynamic deformation of  nuclear surface on the isoscalar dipole modes by using a translation-invariant model of small oscillations of finite Fermi systems based on the Vlasov kinetic equation  \cite{Abr-05}. This semiclassical model explicitly uses the collective variables describing nuclear surface oscillations, which makes it possible to separate the dynamic surface effects in isoscalar dipole modes and is therefore suitable for studying our problem. 
We will focus on the variation of the phase-space distribution function, which is generated by the dipole dynamic deformation of the nuclear surface. This variation allows us to obtain information about the dynamic effects of the nuclear surface in isoscalar dipole excitations, in particular, by using such  physical quantities as the strength function and the velocity field. 

In Section II, the description of the dynamic nuclear surface within our kinetic model is briefly recalled, while in Section III the dynamic-surface approximation to the study the surface effects in the isoscalar dipole modes of nuclei is considered. 
The results of numerical calculations of the strength function and the velocity field for isoscalar dipole modes in the dynamic-surface approximation and the comparison with different approximations are discussed in Section IV. Section V contains the summary. Mathematical details on the isoscalar dipole velocity field are given in the Appendix.

\section{Dynamic nuclear surface in kinetic model}
\label{dynsurf}

In kinetic model of small oscillations of finite Fermi systems with moving surface \cite{Abr-05}, a nucleus is treated as a gas of interacting nucleons confined to a spherical cavity with moving surface. The dipole surface vibrations are described by a set of collective variables $\delta R_{1M}(t)$ according to the equation of surface motion
\bel{eq1}
R(\vartheta ,\varphi ,t)=R+\sum\limits_M {\delta R_{1M} (t)} Y_{1M}(\vartheta ,\varphi ).
\end{equation}
This equation expresses the time-dependent position of the droplet surface $R(\vartheta,\varphi,t)$, which is a sphere with radius $R$ in equilibrium. The variation of the phase-space distribution function $\delta n(\r,\p,t)$ related to the isoscalar dipole excitations is determined by the linearized Vlasov equation 
\bel{eq2}
\frac{\partial }{\partial t}\delta n(\r,\p,t)+\frac{\p}{m}\frac{\partial }{\partial \r}
[ \delta n(\r,\p,t)
-\frac{dn_{0} }{d\epsilon }(\delta V(\r,t)+V_{ext} (\r,t)) ]=0
\end{equation}
and satisfies the boundary conditions at the moving surface that ensure a free moving surface
\bel{eq3}
\left[ {\delta n(\r,\p_{\bot },p_{r}, t)\,-\,\delta n(\r,\p_{\bot },-p_{r}, t)} \right]\vert \,_{r=R}
=-2p_{r} \,\frac{dn_{0} }{d \epsilon }\,\frac{\partial 
}{\partial t}\,\delta R(\vartheta ,\varphi ,t) 
\end{equation}
and 
\bel{eq4}
\delta \Pi_{rr} (\r,t)\vert_{ r=R} \,=\,0.
\end{equation}
Here $p_{r}$ is the radial momentum of the particle, and 
$\p_{\bot }= (0, p_{\vartheta }, p_{\varphi })$. The variation 
of the normal component of the momentum flux tensor 
$\delta\Pi_{rr}(\r,t)$ is defined as \cite{Lifshitz-79} 
\bel{eq5}
\delta \Pi_{rr} (\r,t)=\int d\p\,p_{r} \,v_{r}\,[\delta n(\r,\p,t)-\frac{dn_{0} }{d\epsilon }\,\delta V(\r,t)].
\end{equation}
The equilibrium distribution function $n_{0}(\r,\p)$ depends only on the one-particle energy $\eps(\r,\p)$  and we use the approximation of the Thomas-Fermi type:
\bel{eq6}
\frac{dn_{0} (\epsilon (\r,\p))}{d\epsilon}\,=\,-\frac{4}{h^{3}}\delta (\epsilon -\epsilon_{F} ),
\end{equation}
where $\eps_{F}$ is the Fermi energy. The mirror reflection boundary condition (\ref{eq3}) determines a connection between the variation of the phase-space distribution function and surface oscillations, while the boundary condition (\ref{eq4}) provides the consistent motion of particles inside the nucleus and the free surface. 

The residual interaction between nucleons is taking into account in the kinetic equation (\ref{eq2}) and in the boundary condition (\ref{eq4}) due to the mean-field variation 
$\delta V(\r,t)$ given by 
\bel{eq7}
\delta V(\r,t)=\int d{\r}'d\p\, v(\r,{\r}')\,\delta n({\r}',\p,t).
\end{equation}
In order to obtain an explicit solution of the linearized Vlasov equation, we assume a separable effective interaction of the form 
\bel{eq8}
v (\r,{\r}')=\kappa_{1} \sum\limits_M{ r\,{r}'Y_{1M}(\vartheta ,\varphi)\,Y_{1M}^{\ast}({\vartheta}',{\varphi}')} ,
\end{equation}
where $\kappa_{1}$ is the isoscalar dipole interaction strength.

Taking into account that isoscalar dipole excitations of finite Fermi systems are a second-order effect with respect to the dipole moment, we assume that collective isoscalar dipole modes are excited by a weak external field of the type
\bel{eq9}
V_{ext} (\r,t)=\beta\, \delta (t)\,Q^{(3)}(r)\,Y_{10}(\vartheta,\varphi ),
\end{equation}
where $Q^{(3)}(r)=r^3$ is the second-order dipole moment, $\delta (t)$ is the Dirac delta-function in time, and $\beta$ is a parameter that describes the external field strength.

To find the solution to dynamic equations (\ref{eq1}) and (\ref{eq2}) with boundary conditions (\ref{eq3}) and (\ref{eq4}), it is reasonable to change from phase variables 
$(\r, \p)$ to new variables $(r, \eps, l, \alpha, \beta, \gamma)$. Here, $\eps$ is the particle energy, $l$ its angular momentum, and $\alpha, \beta, \gamma$ are the Euler angles, which describe the rotation to the coordinate system with the $z$-axis directed along the vector $\bf l=\r \times \p$ and the $y$-axis directed 
along the vector $\r$. The (Fourier transformed in time) 
variation of the distribution function can be written in terms of new variables as the series expansion \cite{Brink-86} 
\belar{eq10}
\delta n(\r,\p,\omega)&=&\sum\limits_{MNL} \,[\delta 
n_{MN}^{L+} (\epsilon ,l,r,\omega)+\delta n_{MN}^{L-} (\epsilon ,l,r,\omega )]\nonumber \\
&\times&(D_{MN}^{L} (\alpha ,\beta ,\gamma ))^{\ast }\,Y_{LN} (\frac{\pi }{2},\frac{\pi }{2}) ,
\end{eqnarray}
where $L=1$ and $M=0$ for the variation of the distribution function related to the isoscalar dipole excitations induced by the external field (\ref{eq9}), the quantities $D_{MN}^{L} (\alpha ,\beta ,\gamma )$ are the Wigner rotation matrices and the value of $\omega $ defines the excitation energy of the nucleus ($E=\hbar \omega $). The functions $\delta n_{MN}^{L\pm}$ refer to particles having positive ($\delta n_{MN}^{L+}$) or negative ($\delta n_{MN}^{L-}$) components of the radial momentum $p_r$.

\section{Dynamic-surface approximation}
\label{dyn-approx}
Taking into account the boundary condition (\ref{eq3}) the solution of the linearized Vlasov equation (\ref{eq2}) for the (Fourier transformed in time) variation of the phase-space distribution function can be written as
\bel{eq11}
\delta n(\r,\p,\omega)\,=\,\delta n_{stat}(\r,\p,\omega)\,+\,\delta n_{dyn}(\r,\p,\omega).
\end{equation}
Here $\delta n_{stat}(\r,\p,\omega)$ is the solution for a system confined by the static (fixed) surface while the variation $\delta n_{dyn}(\r,\p,\omega)\sim \delta R_{1M}(\omega )$ is due to surface oscillations. 
Taking into account the expansion (\ref{eq10}), the solution (\ref{eq11}) can be rewritten in terms of new variables in the form 
\bel{eq12}
\delta n_{N}^{\pm } (r,\epsilon ,l,\omega )=\,\delta n_{stat,N}^{\pm } (r,\epsilon ,l,\omega )\,+\,\delta n_{dyn,N}^{\pm } (r,\epsilon ,l,\omega ).
\end{equation}
We can find an explicit expression for the distribution function associated with  isoscalar dipole excitations in the presentation (\ref{eq12}) using the equation (106) of paper \cite{Abr-05} with L=1 and M=0\footnote{Note that, instead of $\delta {\tilde {f}}_{0N}^{1\pm}(r,\epsilon ,l,\omega )$, here the notation $\delta n_{N}^{\pm}(r,\epsilon ,l,\omega )$ is used.} 
and then  obtain, in different approximations,  the isoscalar dipole response function, as well as the local dynamic quantities associated with isoscalar dipole modes, in particular, the velocity field .

To evaluate the effect of nuclear surface oscillations on the isoscalar dipole modes we neglect the static-surface contribution $\delta n_{stat}(\r,\p,\omega)$ to the solution 
$\delta n (\r,\p,\omega)$ and consider the approximation (the dynamic-surface approximation) defined as
\bel{eq13}
\delta n(\r,\p,\omega)\,\approx \,\delta n_{dyn}(\r,\p,\omega),
\end{equation}
or in terms of new variables 
\bel{eq14}
\delta n_{N}^{\pm } (r,\epsilon,l,\omega )\,\approx \,\delta n_{dyn,N}^{\pm }(r,\epsilon,l,\omega ).
\end{equation}
We use this approximation in the energy region 10 - 30 MeV where the isoscalar dipole modes are observed in heavy spherical nuclei \cite{Clark-01,Uchida-03,Clark-04,Uchida-04}.

In  kinetic model, the isoscalar dipole response function for a system with a moving surface  is defined as \cite{Abr-02,Abr-16}
\bel{eq21}
R (\omega )=\beta^{-1} \int \,d\r\,r^{3}\,Y_{1M}^{\ast } 
(\vartheta ,\varphi )\,\delta {\bar{{\varrho }}}
(\r ,\omega ). 
\end{equation}
Here the modified particle density variation $\delta {\bar{{\varrho }}}(\r ,\omega )$ is given by
\bel{eq22}
\delta \mathop {\bar{{\varrho }}}(\r,\omega )=\delta \varrho 
(\r,\omega )+\varrho_{0} \,\delta (r-R)\,\delta R_{1M} (\vartheta 
,\varphi ,\omega ),
\end{equation}
where
\bel{eq23}
\delta \varrho (\r ,\omega )=\int \,d\p \,\delta n (\r ,\p ,\omega ).
\end{equation}
In our translation-invariant model the strength associated with the center of mass motion is concentrated at the zero energy. Therefore the response function (\ref{eq21}) can be written in the form 
\bel{eq24}
R(\omega )=R_{intr} (\omega)- R_{c.m.}(\omega ).
\end{equation}
Here the response of the center of mass $R_{c.m}(\omega)$ induced by the external field (\ref{eq9}) is proportional to $1/{{\omega}^{2}}$. It does not excite the system at ${\omega}\neq 0$. Taking into account the expantion of the solution given by the expression (\ref{eq11}), the internal response function $R_{intr} (\omega)$ can be written in the form
\bel{eq25}
R_{intr}(\omega) = R_{stat} (\omega )+ R_{dyn}(\omega )- R_{c.m.} (\omega ),
\end{equation}
where $R_{stat} (\omega)$  is the response function of the system with a fixed surface and the second term $R_{dyn}(\omega)$ is the contribution to the internal response function generated by the moving surface. Below we will consider the internal isoscalar dipole strength function that is determined as
\bel{eq26}
S(E)=-\frac{1}{\pi }Im R_{intr} (E),
\end{equation}
where $E= \hbar \omega $. By using the expansion (\ref{eq25}) we can evaluate the strength function in different approximations, in particular, in the dynamic-surface approximation.

Next, we consider the velocity field associated with the isoscalar dipole modes. This local dynamic quantity describes the spatial distribution of the average nucleon velocity under collective excitation and gives the information about the nature of this excitation. In kinetic theory, the velocity field is determined as 
\bel{eq15}
{\bf u}(\r,\omega )=(m\rho_{0})^{-1}\int {d\p \,\p \,\delta n (\r,\p,\omega)} ,
\end{equation}
where $\rho_{0}$ is the nuclear equilibrium density. We will consider the velocity field in the meridian plane $XZ$ that usually exploited in the random-phase-approximation (RPA) calculations \cite{Hama-99}. In this representation the particle radius-vector is given by $\r =(x, y=0, z)$ or $\r=(r,\vartheta, \varphi =0)$ in the spherical coordinates and the velocity field (\ref{eq15}) can be written as 
\bel{eq16}
{\bf u}(r,\vartheta,\varphi=0,\omega)=u_{r} (r,\vartheta ,\omega ){\bf e}_{r} + u_{\vartheta } (r,\vartheta ,\omega )
{\bf e}_{\vartheta } ,
\end{equation}
where $u_{r}(r,\vartheta ,\omega )$ and $u_{\vartheta }(r,\vartheta , \omega )$ 
are the radial and tangential components of the velocity field vector and ${\bf e}_{r} $, ${\bf e}_{\vartheta } $
are the corresponding unit vectors of the spherical coordinate system. 
The cartesian componens  $u_{x}(r, \vartheta ,\omega )$ and $u_{z}(r, \vartheta , \omega )$  of the velocity field vector (\ref{eq16}) are given in
Appendix.

The expressions for the functions $u_{r}(r, \vartheta ,\omega )$ and $u_{\vartheta  }(r, \vartheta , \omega )$ can be found in the form (see Appendix) 
\bel{eq17}
u_{r} (r,\vartheta ,\omega )=Y_{10} (\vartheta ,0)u_{r} (r,\omega ),
\end{equation}
\bel{eq18}
u_{\vartheta } (r,\vartheta ,\omega )=Y_{11} (\vartheta ,0)u_{\vartheta } (r,\omega ).
\end{equation}
Here the dipole spherical harmonics $ Y_{1m} (\vartheta,0)$ determine the angular dependence of the velocity field, while the radial functions 
$u_{r}(r,\omega )$ and $u_{\vartheta }(r,\omega )$ can be written as
\bel{eq19}
u_{r} (r,\omega)=\pi \frac{1}{\rho_{0} }\frac{1}{r^{2}}\int {d\epsilon } \int {dl} l
\sum\limits_{N=-1}^1 {[\delta n_{N}^{+} (r,\epsilon ,l,\omega )-\delta n_{N}^{-} (r,\epsilon ,l,\omega )]}, 
\end{equation}
\belar{eq20}
u_{\vartheta } (r,\omega )&=&-\frac{i}{\sqrt 2 }\pi \frac{1}{\rho_{0} 
}\frac{1}{r^{2}}\int {d\epsilon } \int {dl} l 
\frac{p_{\bot } (r,l)}{p_{r} (r,\epsilon ,l)\,}\\
&\times&\sum\limits_{N=-1}^1{N[\delta n_{N}^{+} (r,\epsilon ,l,\omega )+\delta n_{N}^{-} (r,\epsilon ,l,\omega )]}, \nonumber
\end{eqnarray}
where $p_{r}(r,\eps,l)=[2m \eps-{(l/r)}^{2}]^{1/2}$ and 
$p_{\bot }(r,l)=l/r$ are radial and tangential momentum of the particle, respectively. 
By means of the expansion (\ref{eq12}), we get the expression of the  velocity field (\ref{eq16}), which allows for the study of the velocity field in various approximations, in particular, in the dynamic-surface approximation. 

\section{Surface effects in strength function and velocity field}
\label{surfeff}
The isoscalar dipole strength function (\ref{eq26}) for heavy spherical nuclei in the energy region of the isoscalar dipole resonances (10-30 MeV) is shown in Fig. 1 in different approximations.
We consider the response of a sample `nucleus' of A=208 nucleons. Dipole strength functions calculated for other values of A, corresponding to other medium-heavy spherical nuclei, are qualitatively similar to the case shown in Fig. 1, in particular, 
in Ref. \cite{Abr-02},  see Fig. 4 (solid curves), it was shown that the isoscalar dipole strength functions for the spherical nuclei ${^{208}Pb}$, ${^{116}Sn}$, and ${^{90}Zn}$ have a similar form, but shift to higher energies with decreasing of A. 
The strength parameter of the isoscalar dipole residual interaction (\ref{eq8}), chosen in order to reproduce the experimental value of the giant monopole resonance 
energy in ${^{208}Pb}$ within our kinetic model \cite{Abr-02}, is $\kappa_{1}=-7.5 \cdot 10^{-3}$ MeV/fm$^{2}$. The corresponding value of the incompressibility modulus equals K$_{A} =$160 MeV. The numerical calculations were carried out using the following values for the nuclear parameters: 
r$_{0}=$ 1.25 fm, $\eps_{F}=$ 30.94 MeV, and m $=$ 1.04 MeV 
(10$^{-22}$sec)$^{2}$/fm$^{2}$.
\begin{figure}[htb]
	\centering
	\includegraphics[width=0.8\textwidth]{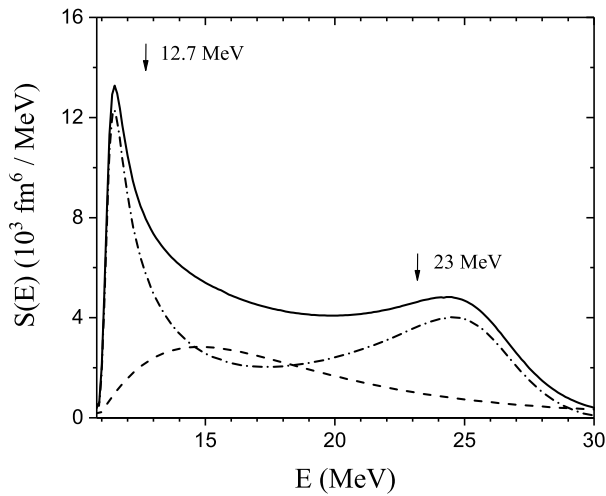}
	\caption{Isoscalar dipole strength function in the energy region of the isoscalar dipole resonances is shown in different approximations: the exact strength function within our model (solid curve), the strength function in the dynamic-surface approximation, see Eq. (\ref{eq13}), (dot-dashed  curve) and evaluated with static-surface boundary condition (dotted curve).The system contains A$=$208 nucleons. The experimental centroid energies of low- and high-energy isoscalar dipole resonances for ${^{208}Pb}$ of Ref. \cite{Uchida-03} are indicated by arrows.}
	\label{fig1}
\end{figure}

By comparing the three curves shown in Fig. 1, we can evaluate the role of  dynamic surface effects to the formation of isoscalar dipole resonances.
The solid curve shows the exact internal strength function within our kinetic model $R_{intr} (\omega)$, which is calculated using the total variation of the distribution function $\delta n_{stat}(\r,\p,\omega)$ induced by the external field (\ref{eq9}), see (\ref{eq25}), (\ref{eq11}). It can be seen from Fig. 1 that the exact strength function has a two-resonance structure. The centroid energies of the low- and  high-energy resonances equal 11.5 and 24.3 MeV, respectively, and are in reasonable agreement with the experimental centroid energies for the isoscalar dipole resonances in the ${^{208}Pb}$ nucleus \cite{Uchida-03}, which are indicated by arrows in Fig. 1. 
The results of  calculations of the internal strength function in the dynamic-surface approximation are shown in Fig. 1 by the dot-dashed curve. We can see that, in the dynamic-surface approximation, the strength function also has a two-resonance structure and, moreover, the centroid energies of the low- and  high-energy resonances are very close to the corresponding centroid energies of the exact strength function. Finally, the dashed curve in Fig.1 shows that the strength distribution in the static-surface approximation does not have a significant effect on isoscalar dipole resonances. 
Thus, both low-energy and high-energy isoscalar dipole resonances are formed due to dynamic surface effects.  
\begin{figure}[htb]
	\centering
	\includegraphics[width=0.7\textwidth]{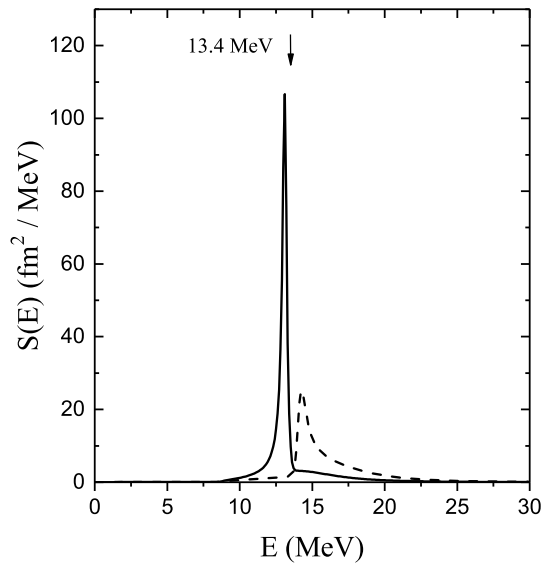}
	\caption{Isovector dipole strength function taken from Ref. \cite{Abr-03} is shown in different approximations: the strength function  with taken into account dynamic surface effects (solid curve), and the one evaluated with static-surface boundary condition (dashed curve).The system contains A$=$208 nucleons. The experimental centroid energy of isovector giant dipole resonance for ${^{208}Pb}$ of Ref. \cite{Woude} is indicated by arrow.}
	\label{fig2}
\end{figure}

\begin{figure*}[!htb]
	\centering
	\includegraphics*[width=0.99\textwidth]{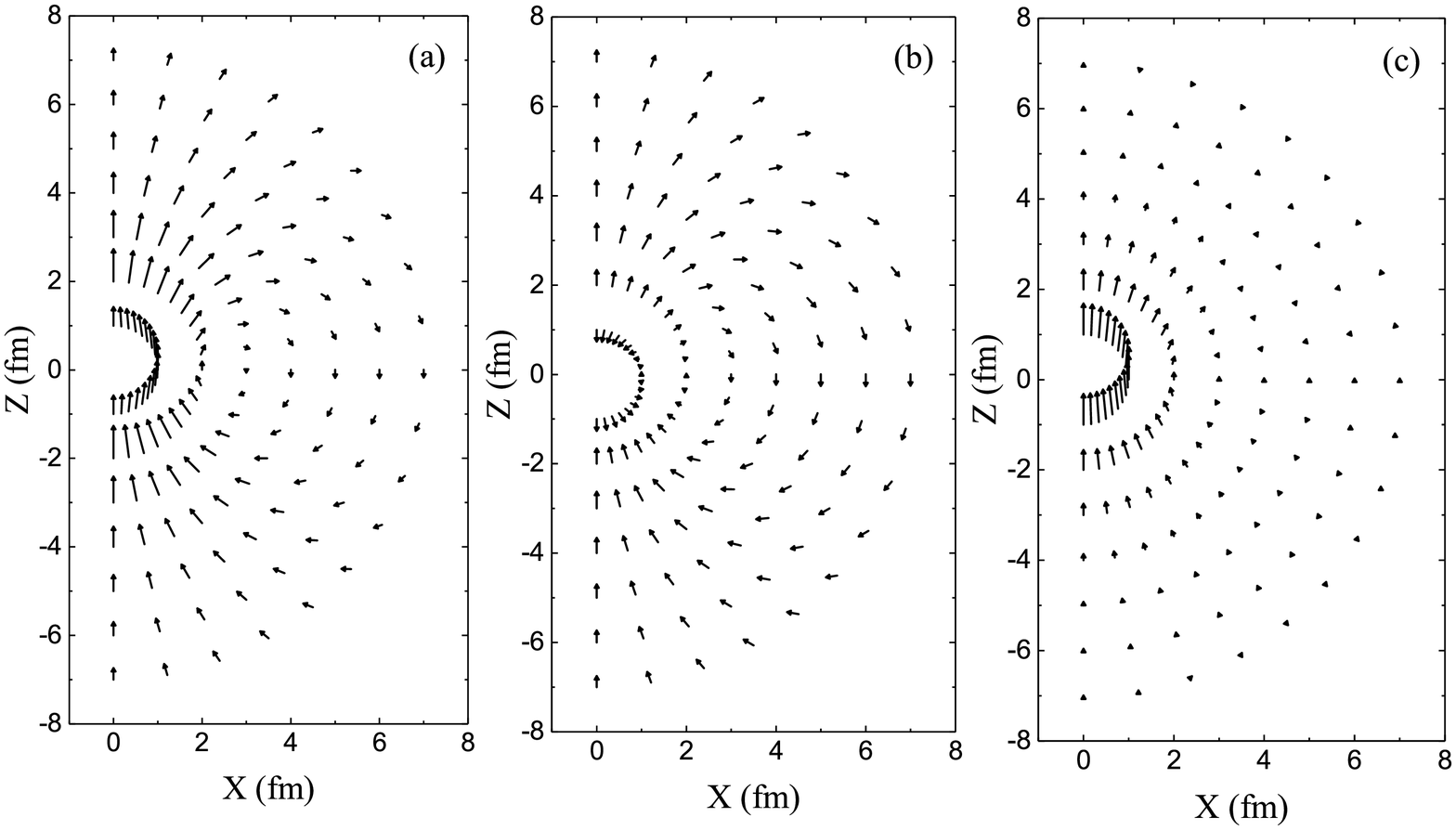}
	\caption{Velocity fields in the $XZ$-plane associated with the low-energy resonance of the isoscalar dipole strength function at the centroid energy 11.5 MeV, see Fig. 1 (solid curve), are shown in different approximations, see Eqs.(\ref{eq12}), (\ref{eq13}): the velocity field evaluated using the exact solution, Fig. 3(a), in the dynamic-surface approximation, Fig. 3(b), and in the static-surface approximation, Fig. 3(c). The system contains A$=$ 208 nucleons.}
	\label{fig3}
\end{figure*}

\begin{figure*}[!htb]
	\centering
	\includegraphics*[width=0.99\textwidth]{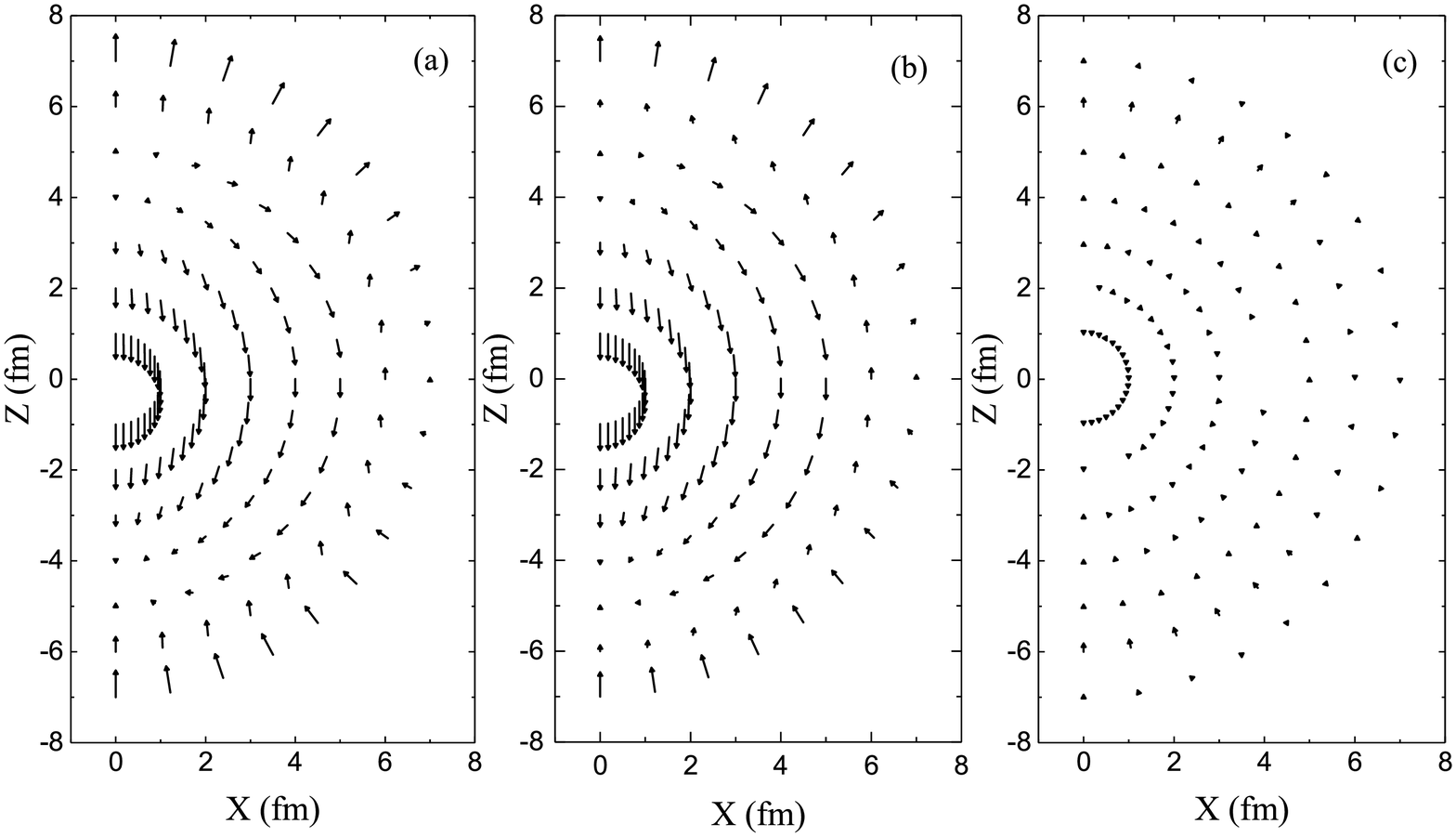}
	\caption{Velocity fields in the $XZ$-plane associated with the high-energy resonance of the isoscalar dipole strength function at the centroid energy of 24.3 MeV, see solid curve in Fig.1, in the same approximations as in Fig. 3. }
	\label{fig4}
\end{figure*}

It is interesting to compare the effect of dynamic surface on  isoscalar dipole resonances with the effect on  the isovector giant dipole resonance.  In Fig. 2 taken from  Ref. \cite{Abr-03}, it can be seen that, in contrast to isoscalar dipole resonances,   dynamic surface effects  lead only to a shift of the isovector giant  dipole resonance to  lower energies , while this resonance is formed  in a Fermi system with a fixed surface. 

To evaluate the effect of the dynamic surface on the character of the velocity fields associated with isoscalar dipole resonances,   in Figs. 3 and 4   are shown in different approximations the velocity fields at centroid energies of the low- and high-energy resonances of the  strength function.  
In Figs. 3(a) and 4(a), the exact velocity fields within our kinetic model, see Eqs. (\ref{eq11}),(\ref{eq15}), are shown. As can be seen, the exact velocity field associated with the low-energy resonance at the centroid energy 11.5 MeV displays a purely vortex (toroidal) character, see Fig. 3(a), while the exact velocity field associated with the high-energy resonance at the centroid energy 24.3 MeV has a compression character, see Fig. 4(a). These results are in qualitative agreement with the previous results of relevant RPA calculations \cite{Vret-00,Vret-02,Kvasil-13,Kvasil-14}.

In Figs. 3(b) and 4(b), the results of calculations of the velocity fields for the isoscalar dipole resonances in the dynamic-surface approximation are shown. Fig. 3(b) shows that, in the dynamic-surface approximation, the velocity field associated with the low-energy resonance has essentially a vortex character similar to the exact velocity field, cf. Figs. 3(b) and 3(a), but this velocity field also involves a compression inside the nuclear fluid. 
On the contrary, in the static-surface approximation, the velocity field for the low-energy resonance  has no vortex motion, see Fig. 3(c), however it has a visible strengthening  inside the nuclear fluid associated with single-particle excitations, which ensures a purely vortex character of the exact velocity field. Therefore, the vortex nature of the low-energy isoscalar dipole resonance is associated with both dynamic surface effects and single-particle excitations inside the nuclear fluid. 
In Fig. 4(b) the velocity field associated with the high-energy isoscalar dipole mode is shown in the dynamic-surface approximation. It can be seen that, in the dynamic-surface approximation, this velocity field has a compression character. Moreover, it looks very similar to the exact velocity field, cf. Figs. 4(b) and 4(a), while in the static-surface approximation, the velocity field for the high-energy isoscalar dipole mode is negligible, see Fig. 4(c). 
Thus, the features of the velocity field associated with the high-energy isoscalar dipole mode are mainly determined by dynamic surface effects .

\section{Summary}
\label{summa}
We have studied the effect of the nuclear surface oscillations on the isoscalar dipole resonances in heavy spherical nuclei within the kinetic model of small vibrations of finite Fermi systems with a moving surface. The isoscalar dipole strength function and the velocity fields associated 
with isoscalar dipole modes were evaluated in different approximations, in particular, in the dynamic-surface approximation, which takes into account only the variation of the phase-space distribution function caused by the moving surface.

It is found that, in the energy region of the isoscalar dipole resonances (10-30 MeV), the strength distribution is mostly determined by dynamic surface effects. The isoscalar strength function of the dipole, evaluated in the dynamical surface approximation, is very similar to the strength function , which is exact within our model.
Moreover, without taking into account dynamic effects of the nuclear surface, both the low-energy isoscalar dipole resonance (the vortex mode) and the high-energy one (the compression mode) cannot be created.

Dynamic surface effects also determine the character of the velocity fields associated with the isoscalar dipole resonances. 
In the dynamic-surface approximation, the velocity field for low-energy resonance  has a vortex character.  At the same time, the velocity field associated with high-energy resonance has a compression character that is approximately the same as  the  velocity field, which is exact within our model. 

The isoscalar dipole resonances (vortex and compression modes), which are observed in heavy spherical nuclei 
are essentially related to dynamic effects of the nuclear surface.
Moreover, the formation of vortex motion in low-energy isoscalar dipole resonance is due to both dynamic surface effects and single-particle excitations inside a nuclear liquid. 

For the further investigation of the nature of collective excitations in nuclei, it would be interesting to study local dynamic quantities associated with changes in  momentum space, in particular, the momentum flux tensor, which contains information about dissipative processes (viscosity, thermal 
conductivity) in a nuclear quantum fluid.

{\bf Acknowledgments.} 
This work is supported in part by the budget program "Support for the development of priority areas of scientific researches" of National Academy of Sciences of Ukraine, project No. 0122U000848.
\appendix
\section{Isoscalar dipole velocity field}
\label{append}
In this Appendix we give some details of the derivation of the analytical expressions (\ref{eq17}) and (\ref{eq18}) for the isoscalar dipole velocity field.

Taking the $z$-axis in the direction of the external field, we are interested in the velocity field (\ref{eq15}) in the meridian plane $XZ$. First we find the cartesian coordinates of the velocity field $u_{x}({\bf r},\omega)$ and $u_{z}({\bf r},\omega)$:
\bel{comp}
u_{x(z)}({\bf r},\omega)=\frac{1}{m\varrho_{0}}\int
d{\bf p}p_{x(z)}\delta n({\bf r},{\bf p},\omega)\,.
\end{equation}
It is useful to express them through the spherical components $u_{1m}({\bf r},\omega)$:
\belar{uxz}
&&u_{x}({\bf r},\omega)=-\frac{1}{\sqrt{2}}[u_{11}({\bf r},\omega)-u_{1-1}({\bf r},\omega)],\nonumber\\
&&u_{z}({\bf r},\omega)=u_{10}({\bf r},\omega),
\end{eqnarray}
that are related to the corresponding components of the momentum in
the lab system,which can be written as
\bel{pm}
p_{1m}(\vartheta_{p},\varphi_{p})=pC_{1m}(\vartheta_{p},\varphi_{p})\,,
\end{equation}
with $p=\sqrt{2m\epsilon}$ and $C_{1m}(\vartheta_{p},\varphi_{p})=\sqrt{\frac{4\pi}{3}}Y_{1m}(\vartheta_{p},\varphi_{p})$.

Now we make the same rotation of the coordinate system as in the expansion (\ref{eq10}). 
In the rotated system, $p'_{1m}(\vartheta'_{p},\varphi'_{p})=pC_{1m}(\vartheta'_{p},\varphi'_{p})$.
Note that $\vartheta'_{p}= \frac{\pi}{2}$, while $\frac{\pi}{2}\leq
\varphi'_{p}\leq 3\frac{\pi}{2}$.
The spherical components of the momentum in the rotated system are conveniently expressed in terms of the cartesian coordinates:
\belar{p1m}
&&p_{11}(\vartheta'_{p},\varphi'_{p})=-\frac{1}{\sqrt{2}}(p_{x'}+ip_{y'}),\nonumber\\
&&p_{10}(\vartheta'_{p},\varphi'_{p})=p_{z'},\nonumber\\
&&p_{1-1}(\vartheta'_{p},\varphi'_{p})=\frac{1}{\sqrt{2}}(p_{x'}-ip_{y'})\,,
\end{eqnarray}
which, for given $\epsilon$ and $l$, are given in terms of the magnitude of the radial $p_r$ and tangential $p_{\bot}$ components of the particle momentum:
\belar{pxyz}
&&p_{z'}=0,\nonumber\\
&&p_{y'}=\pm p_{r}=\pm \sqrt{2m\epsilon-\frac{l^{2}}{r^{2}}},\nonumber\\
&&p_{x'}=-p_{\bot}=-\frac{l}{r}\,.
\end{eqnarray}

According to Eq. (2.24) of \cite{bs} the spherical components  of the particle momentum in the two reference frames are related by
\bel{tran}
p_{1m}(\vartheta_{p},\varphi_{p})=\sum_{n}{\Big (}{\cal
D}^{1}_{mn}(\alpha,\beta,\gamma){\Big
)}^{*}p_{1n}(\vartheta'_{p},\varphi'_{p})\,.
\end{equation}
Now we consider the  velocity field \req{eq15}, written as
\bel{umod}
{\bf u}({\bf r},\omega)=\frac{1}{m\varrho_{0}}\int d{\bf r}''\int
d{\bf p} {\bf p} \delta ({\bf r}-{\bf r}'')
\delta n({\bf r}'',{\bf p},\omega).
\end{equation}
We change the integration variables from phase variables $({\r}'', \p)$ to new variables $(r, \eps, l, \alpha, \beta, \gamma)$ according to \cite{Brink-86}
\bel{Jacobian}
\int d{\bf r}''\int d{\bf p}=
\int dr''\int d\epsilon\int
d
\frac{l}{v_{r}}\int_{0}^{2\pi}d
\int_{0}^{\pi}\sin\beta
d\beta\int_{0}^{2\pi}d\gamma,
\end{equation}
take the usual multipole expansion of the $\delta$-function
\bel{dexp}
\delta ({\bf r}-{\bf r}'')=\frac{\delta(r-r'')}{r^{2}}\sum
_{L'M'}Y_{L'M'}(\vartheta,\varphi)Y_{L'M'}^{*}(\vartheta'',\varphi'')\,,
\end{equation}
and the transformation analogous to Eq. \req{tran}
\bel{ylm}
Y_{L'M'}^{*}(\vartheta'',\varphi'')=\sum_{N'}{\cal
D}^{L'}_{M'N'}(\alpha,\beta,\gamma)
Y_{L'N'}^{*}(\frac{\pi}{2},\frac{\pi}{2})\,.
\end{equation}
Moreover, we use the expansion \req{eq10} for given  $L$ and $M=0$.
 
Then the spherical components of the velocity field \req{umod} are given by:
\belar{u1m}
u_{1m}^{L}({\bf r},\omega)&=&\frac{1}{m\varrho_{0}}\sum_{N}\sum_{L'M'N'}
Y_{L'M'}(\vartheta,\varphi)Y_{L'N'}^{*}(\frac{\pi}{2},\frac{\pi}{2})
Y_{LN}(\frac{\pi}{2},\frac{\pi}{2})
\nonumber\\
&\times& 
\int dr''\frac{\delta(r-r'')}{r^{2}}\int d\epsilon\int
dl\frac{l}{v_{r}}\delta n^{L}_{0N}(\epsilon,l,r'',\omega)\nonumber\\
&\times&
\sum_{n}p_{1n}(\epsilon,l,r'')
\int_{0}^{2\pi}d\alpha\int_{0}^{\pi}d\beta\sin\beta
\int_{0}^{2\pi}d\gamma\nonumber\\
&\times&{\cal D}^{L'}_{M'N'}(\alpha,\beta,\gamma)
{\Big (}{\cal D}^{1}_{mn}(\alpha,\beta,\gamma)
){\Big )}^{*}
{\Big (}{\cal D}^{L}_{0N}(\alpha ,\beta,\gamma){\Big )}^{*}
\,.
\end{eqnarray}
The angular integrals are given by (see Ref.\cite{bs}, p. 148)
\belar{angint}
&&\int_{0}^{2\pi}d\alpha\int_{0}^{\pi} d\beta\sin\beta
\int_{0}^{2\pi}d\gamma {\cal D}^{L'}_{M'N'}(\alpha,\beta,\gamma)
{\Big (}{\cal D}^{1}_{mn}(\alpha,\beta,\gamma){\Big
)}^{*}{\Big (}{\cal D}^{L}_{0N}(\alpha ,\beta,\gamma){\Big
)}^{*}
\nonumber\\
&&=(-)^{m-n}\frac{8\pi^{2}}{2L+1}<L'1M'-m|L0><L'1N'-n|LN>.
\end{eqnarray}

For given $L$ and $M=0$, the Clebsh-Gordan coefficients vanishes, unless
$|L-1|\leq
L' \leq L+1$, $M'=m$ (this fixes $M'$) and $N'=N+n$. Moreover,
$Y_{L'N'}^{*}(\frac{\pi}{2},\frac{\pi}{2})$ vanishes unless $L'+N'$ is even. The components of the particle momentum $p_{1n}=0$ for $n=0$, thus $n=\pm 1$. This implies $N'=N\pm
1$ and $L'=L\pm 1$ (no term $L'=L$). Therefore, for given $L$ and $M=0$, 
Eq. (\ref {u1m}) gives
\belar{zvelff}
u_{1m}^{L}({\bf r},\omega)&=&
(-)^{m}[<(L-1)1m-m|L0>Y_{L-1m}(\vartheta,\varphi) u_{L,L-1}(r,\omega)
\nonumber\\
&+&
<(L+1)1m-m|L0>Y_{L+1m}(\vartheta,\varphi)u_{L,L+1}(r,\omega)],\,
\end{eqnarray}
with the radial functions
\belar{radfpm}
u_{L,L\pm 1}(r,\omega)&=&-\frac{1}{m\varrho_{0}}\frac{8\pi^2}{2L+1}
\sum_{N=-L}^{L}\sum_{n=\pm1}Y_{LN}(\frac{\pi}{2},\frac{\pi}{2})
Y_{L\pm1N+n}^{*}(\frac{\pi}{2},\frac{\pi}{2})\, 
\nonumber
\\
&\times& <(L\pm 1)1(N+n)-n|LN>
\nonumber\\ 
&\times& 
\frac{1}{r^{2}}\int d\epsilon \int
dl \frac{l}{v_{r}}\delta n^{L}_{0N}(\epsilon,l,r,\omega)p_{1n}(\epsilon,l,r)\,.
\end{eqnarray}

For $L=1$ the equations \req{zvelff} and \req{radfpm} gives the analytical expressions for the spherical components of the dipole velocity field. They can be written as
\belar{m1m}
&&u_{10}^{1}({\bf r},\omega)=
Y_{00}u_{10}(r,\omega)
-\sqrt{\frac{3}{5}}Y_{20}(\vartheta,\varphi)u_{12}(r,\omega),
\nonumber\\
&&u_{1\pm}^{1}({\bf r},\omega)=
{\pm}\sqrt{\frac{3}{10}}Y_{21}(\vartheta,\varphi)u_{12}(r,\omega),
\end{eqnarray}
with the radial functions
\belar{u1rad10}
u_{10}(r,\omega)&=&-i\sqrt{\frac{1}{3}}\pi
\frac{1}{\varrho_{0}}\frac{1}{r^2} \int d\epsilon \int dll
\\
&\times&
\sum_{N=-1}^{1}\{ i[\delta n^{1+}_{0N}(\epsilon,l,r,\omega)-
\delta n^{1-}_{0N}(\epsilon,l,r,\omega)]\nonumber\\
&+&N \frac{1}{p_r(\epsilon,l,r)} \frac{l}{r}
[\delta n^{1+}_{0N}(\epsilon,l,r,\omega)+
\delta n^{1-}_{0N}(\epsilon,l,r,\omega)]\}
\nonumber,
\end{eqnarray}
\belar{u1rad12}
u_{12}(r,\omega)&=&-i\sqrt{\frac{2}{3}}\pi
\frac{1}{\varrho_{0}}\frac{1}{r^2} \int d\epsilon \int dll
\\
&\times&
\sum_{N=-1}^{1}\{- i[\delta n^{1+}_{0N}(\epsilon,l,r,\omega)-
\delta n^{1-}_{0N}(\epsilon,l,r,\omega)]\nonumber\\
&+&\frac{N}{2} \frac{1}{p_r(\epsilon,l,r)} \frac{l}{r}
[\delta n^{1+}_{0N}(\epsilon,l,r,\omega)+
\delta n^{1-}_{0N}(\epsilon,l,r,\omega)]\}
\nonumber.
\end{eqnarray}

Finally, taken into account that in the meridian plane $XZ$ the particle radius-vector is given by  $\r=(r,\vartheta, \varphi =0)$, and the expressions for the radial and tangential components of the velocity field in terms of  the the cartesian components are defined as
\bel{urr}
u_{r}(r,\vartheta,\omega)=
\sin(\vartheta)u_{x}(r,\vartheta,\omega)+
\cos(\vartheta)u_{z}(r,\vartheta,\omega),
\end{equation}
\bel{urt}
u_{\vartheta}(r,\vartheta,\omega)=\cos(\vartheta)u_{x}(r,\vartheta,\omega)-\sin(\vartheta)u_{z}(r,\vartheta,\omega),
\end{equation}
we get the analytical expressions (\ref{eq17}) and (\ref{eq18}) for the isoscalar dipole velocity field.

\end{document}